\documentclass[letterpaper]{article}
\def\year{2021}\relax
\usepackage{aaai21} 
\usepackage{times} 
\usepackage{helvet} 
\usepackage{courier} 
\usepackage[hyphens]{url}
\usepackage{graphicx} 
\usepackage{natbib}  
\usepackage{caption}
\usepackage{tabularx}
\usepackage{tabulary}
\usepackage{algorithm2e}
\usepackage{amsmath}
\title{Belief and Persuasion in Scientific Discourse on Social Media: A Study of the COVID-19 Pandemic}
\author{
    Salwa Alamir,\textsuperscript{\rm 1} 
    Armineh Nourbakhsh,\textsuperscript{\rm 2} Cecilia Tilli,\textsuperscript{\rm 1} 
    Sameena Shah\textsuperscript{\rm 2} 
    Manuela Veloso\textsuperscript{\rm 2}\\
}
\affiliations{
    \textsuperscript{\rm 1}J.P. Morgan AI Research, London, U.K.\\
    \textsuperscript{\rm 2}J.P. Morgan AI Research, New York, U.S.\\
    \{salwa.alamir, armineh.nourbakhsh, cecilia.tilli, sameena.shah, manuela.veloso\}@jpmchase.com
}

\begin{document}

\maketitle

\begin{abstract}
Research into COVID-19 has been rapidly evolving since the onset of the pandemic. This occasionally results in contradictory recommendations by credible sources of scientific opinion, public health authorities, and medical professionals. In this study, we examine whether this has resulted in a lack of trust in scientific opinion, by examining the belief patterns of social media users and their reactions to statements related to scientific facts. We devise models to mine belief and persuasion in Twitter discourse using semi-supervised approaches, and show the relationship between lack of belief and insurgence of paranoia and conspiracy theories. By investigating these belief patterns, we explore the best persuasion tactics for communicating information related to COVID-19.
\end{abstract}

\section{Introduction}
Social media has emerged as a breeding ground for spreading conspiracy theories and misinformation as the global COVID-19 pandemic continues to spread. Furthermore, according to 
Jaiswal et. al,
\cite{jaiswal_2020} the regular issuance of contradictory guidelines by national and global authorities, and the lack of scientific consensus with conclusive evidence, have led to an environment of paranoia, mistrust, and the rapid increase of conspiracy theories. This begs the question, are people losing trust in science? 
\par Previous studies on discourse in social media have focused on the opposition to credible news sources \cite{boberg2020}, the identification of intentional misinformants \cite{schild2020}, and the surfacing of racist stereotyping \cite{ferrara2020}. This paper focuses on some of the top scientific sources present on Twitter. As there is no accepted method of quantifying the measurement of trust \cite{prochazka2019}, we deem a scientific source trustworthy if people believe the tweets they release. After identifying belief, we investigate potential persuasion tactics in order to relay a COVID-related message such that it induces belief. More concretely, we address the following questions: 1) Are people losing trust in science? Does this mistrust breed more controversy? 2) Which persuasion tactics work best when discussing COVID-19 related issues/controversies?

The remaining sections of these paper lay out our methodology and findings. Our contributions include:

\begin{itemize}
    \item A measure of belief, and an analysis of the topics that have the highest/lowest levels of belief, the sources that have the highest/lowers levels of belief, and whether COVID-related tweets contain more/less/similar belief to other tweets by the same source.
    \item An analysis on the median sentence structure that induces belief, and predicting belief based on this structure, such that it can be used in future as a persuasion tactic.
\end{itemize}

\section{Are People Losing Trust in Science?}
In order to quantify trust in science, we obtain tweets from top scientific sources on Twitter, and the responses from the individuals interacting with these scientific sources. By classifying the responses, we are able to ascertain which tweet topics resulted in the highest levels of belief, who posted them, and whether or not COVID-related tweets have more/less/similar belief to other scientific posts.

\subsection{Data Collection}
The first step is to define a collection of user accounts belonging to scientific authorities who are likely to discuss COVID-19 frequently. To do this, we conducted a Twitter search using keywords ``COVID AND study AND according to'', and mined tweets posted between January 1, 2020 and July 1, 2020. The resulting tweets were ranked by their engagement (number of likes, retweets, and responses). The top 100 tweets came from 26 unique users who were targeted for this study and are listed in Table \ref{tab:user_handles}.

\begin{table}
\caption{Table of scientific sources user handles}
\label{tab:user_handles}
\begin{tabular}[ht!]{cccc} 
\hline
cdcgov & JHSPH\_CHS & WHO \\ 
THLresearch & coimmunityproj & UAlberta\_FoMD \\ hpscireland & CovidActNow & harvardmed \\ 
NIH & OxfordMedSci & Cambridge\_Uni \\ 
US\_FDA & bmj\_latest & IDMOD\_ORG \\ 
umnmedschool & NEJM & imperial\_SoM \\ 
HealthNYGov & LSHTM & YaleMed \\ 
UniversityLeeds & StanfordMed & PHE\_uk \\  ScrippsHealth & TAMUmedicine & \\
\hline
\end{tabular}
\end{table}

For each user, we used the Twitter API to request the latest 3,200 tweets posted to their timeline. Due to limitations imposed by Twitter, we were unable to request the responses to these tweets directly. Therefore, we utilized Twitter's search API\footnote{\url{https://developer.twitter.com/en/docs/tweets/timelines/api-reference}} to request tweets that mentioned each user handle. This resulted in up to 100,000 mentions per scientific user. We then identified whether each mention was in fact a response to a tweet posted by the handle. This was possible by filtering through each mention's ``in\_reply\_to\_status\_id\_str" field. The final result was a collection of 4,064 tweet-and-response combinations that we will use going forward.

\subsection{Data Augmentation}
We created a labelled dataset by sampling 4 responses to each tweet. Three Researchers labelled each response on a three-way scale---Yes (reflecting belief), No (reflecting disbelief), or Maybe when uncertain (e.g. in cases where sarcasm could be present). This resulted in 1,000 labeled tweets, of which 892  passed a majority-vote filter and were used in the remainder of this study. Since our labelled dataset was small, we opted to augment the data before training a classifier to detect belief. By replacing every word in every tweet with the 2 most similar words, we were able to augment a subset of our labelled data from 200 tweets to 3,368 tweets for training. To find the most similar words, we used the ``glove-twitter-25'' embeddings\footnote{\url{https://nlp.stanford.edu/projects/glove/}} generated by the GloVe model \cite{Pennington14glove} and selected the top two words with the closest cosine similarity. In the vector space, some words that are in proximity to each other are antonyms rather than synonyms. Thus completing a data augmentation by replacing with an opposite word can negatively affect the model's performance. We overcome this by using VaderSentiment \cite{Deho2018} to perform a sentiment analysis on both the original tweet and the augmented version. VADER is a rule-based and lexicon-based sentiment analyzer designed for social media posts. It assigns a compound score between -1 (negative sentiment) and 1 (positive sentiment) to each message. The augmented tweets were filtered to those that had a change of score within 0.05 of the original compound value. 

\begin{table}
\caption{Examples of Augmented Tweets}
\label{tab:augmented_text}
\begin{tabulary}{\linewidth}{LL} 
\hline
Original Text & Augmented Text \\
\hline

It is hard to \textbf{keep} up with the guide lines. It's a moving target. 
& It is hard to \textbf{stay} up with the guide lines. 
\\ 
So why are we changing the guidelines to only \textbf{test} symptomatic people? & So why are we changing the guidelines to only \textbf{exam} symptomatic people? \\ 
\hline
\end{tabulary}
\end{table}

\subsection{Classifying Belief}
\par We applied pre-processing techniques to clean each tweets by lower-casing the text, removing special symbols such as the ``@'' symbols and hashtags (``\#''), removing punctuation, URLs and stop words (via NLTK\footnote{\url{https://www.nltk.org/}}). We then tokenized and lemmatized each tweets. Next, we classified belief using three models, a traditional Support Vector Machines (SVM) \cite{Dilrukshi2013}, a Random Forest model using weighted vectors, and a Long-Short-Term-Memory (LSTM) Network \cite{Li2016}. The hyperparameters used for each model follow: \textbf{LSTM}: Using Keras with the Tensorflow backend, we apply a 50\% dropout, 20\% recurrent dropout, categorical cross-entropy loss and Adam optimizer. Final layer has softmax function for multi-class classification. We run 10 epochs with batch size 10. \textbf{RandomForest}: We use Tfidf vectorization via sci-kit learn after the pre-processing step, as described above. 1,000 estimators are implemented for ensemble learning. \textbf{Linear SVM}: We use Linear Support Vector Classifier (SVC) via sci-kit learn in order to classify the data based on a best-fit hyperplane. We use a Tfidf vectorizer in addition to n-grams ranging from n=1 to n=3. The text is cleaned as before.
 
Many tweets posed challenges to detecting belief, especially in the presence of sarcasm. To opt for more reliable predictions, we removed the ``Maybe'' category from the labels. This resulted in 565 labeled tweets. Table \ref{tab:classification} displays the final results. Based on the results we proceeded to utilize the LSTM model to predict belief in the remaining 3,064 unlabeled tweets. For the remainder of this study, these labels are used to analyze belief patterns.

\begin{table}
\centering
\caption{Classification of belief in tweets.}
\label{tab:classification}
\begin{tabular}[ht!]{cccc} 
\hline
Method & Accuracy & Precision & Recall \\
\hline
 LSTM & 77.19\% & 0.74 & 0.73\\ 
 RandomForest & 69.91\% & 0.70 & 0.70 \\ 
 Linear SVM & 73.45\% & 0.74 & 0.73\\ 
\hline
\end{tabular}
\end{table}

\subsection{Identifying Common Clusters of Discussion}
In order to identrify how conversations cluster around topics, we follow the process outlined in \cite{nourbakhsh}. Using hashtags, we construct a graph where each node is a user and each edge identifies whether the adjacent users have posted a tweet using the same hashtag. The edges are weighted, in that they reflect the number of tweets with common hashtags between each pair of users.

\begin{figure}[h]
    \centering
    \includegraphics[width=\linewidth]{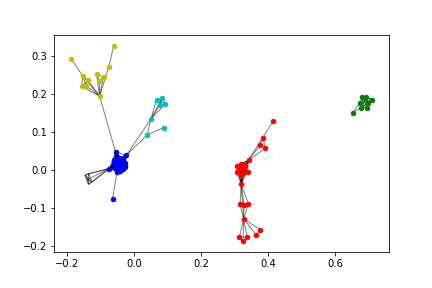}
    \caption{Hashtag-based clusters in COVID-19 tweets.}
    \label{fig:louvain}
\end{figure}

Given this weighted, undirected graph, we identify communities by applying the Louvain Community Detection algorithm\footnote{\url{https://python-louvain.readthedocs.io/en/latest/}}, which performs heirarchical clustering with the aim of maximizing a modularity measure. The resulting five communities are shown in Figure \ref{fig:louvain}. The different communities are highlighted in different colors. They are plotted on the chart using the Networkx library\footnote{\url{https://networkx.org/documentation/stable/index.html}} and laid out according to the Fruchterman-Reingold force-directed algorithm \cite{Fruchterman1991}. 

\subsection{Profiling Belief in Each Cluster}
We identify prominent topics by inspecting the top hashtags for each community. Calculating the percentage of tweets classified as belief (``Yes'') indicates the level of belief present in each community. Table \ref{tab:communities} shows the communities with their corresponding hashtags and belief ratio. Some of the communities have 0\% of the tweets classified as positive belief. This is due to our definition of belief based on responses. Retweets are not incorporated as they remain ambiguous, with some users retweeting to show support, and others as a form of mockery. Thus in our analysis of responses, we find that tweets that contain hashtags such as ``stopmaskinghealthypeople'' and ``worldhoaxorganisation'' show no belief. Other prominent hashtags in the corpus included ``scamdemic'' and ``plandemic'', reinforcing the presence of conspiracy theories. This was consistent with our expectation that discussions involving mistrust and conspiracy theories had tweets labeled with little (or no) belief compared to discussions with no such indication, such as community 5, with ``mentalhealthmatters'' and ``loveisessential''. 

\begin{center}
\begin{table*}[htbp]
\caption{Discussion communities with top hashtags and percentage of belief.}
\label{tab:communities}
\begin{tabular}{ clc }
\hline
Community & Hashtags & Percent Belief \\
\hline
 1 & \begin{tabular}[c]{@{}l@{}}coronavirus, COVID, hqcworks, athome, azithromyin, backtoschool,cdc,coronadebat\end{tabular} & 0\%\\ 
 2 & \begin{tabular}[c]{@{}l@{}}COVID19, prophylactic, COVID, millionaires, health, nhs, vaccination,virus,who\end{tabular} & 40\%\\ 
 3 & \begin{tabular}[c]{@{}l@{}}billgatesbioterrorist, worldhoaxorganisation,defundthewho, faucithefraud,\end{tabular} &0\%\\
 & hydroxycholoroquineworks & 0\%\\ 
 4 & stopmaskingchildren, stopmaskinghealthypeople, sixpercent   &0\% \\ 
 5 & loveisnottourism, loveisessential,mentalhealthmatters & 85 \%\\ 
\hline
\end{tabular}
\end{table*}
\end{center}

\subsection{Belief in COVID-related discussions}
Not all of the updates tweeted by the scientific sources were COVID-related. We therefore compare the percentage of belief in tweets that contain the word ``COVID" or ``Corona" with those that do not. The findings are shown in the Table \ref{tab:COVIDbelief}. We find that the scientific sources have more messages about COVID in general compared to other scientific updates. It is also apparent that belief is not high in general; both groups show that less than half the responses exhibit belief in the original message. Belief is also particularly low with regards to COVID-related tweets, even for the same user. Non-COVID related tweets have more ``Yes" responses relative to total responses; 1.9 times more than COVID-related tweets.

\begin{center}
\begin{table}
\caption{Belief in COVID vs. Non-COVID Tweets}
\label{tab:COVIDbelief}
\begin{tabular}[ht!]{cccc} 
\hline
 Tweet Contains & Yes & No & Percent\\
\hline
 COVID & 473 & 2546 & 15.7\%\\ 
 Non-COVID & 315 & 730 & 30.0\% \\ 
\hline
\end{tabular}
\end{table}
\end{center}

\subsection{Belief in Scientific Sources}
\begin{figure}[tph!]
\centerline{\includegraphics[width=\linewidth]{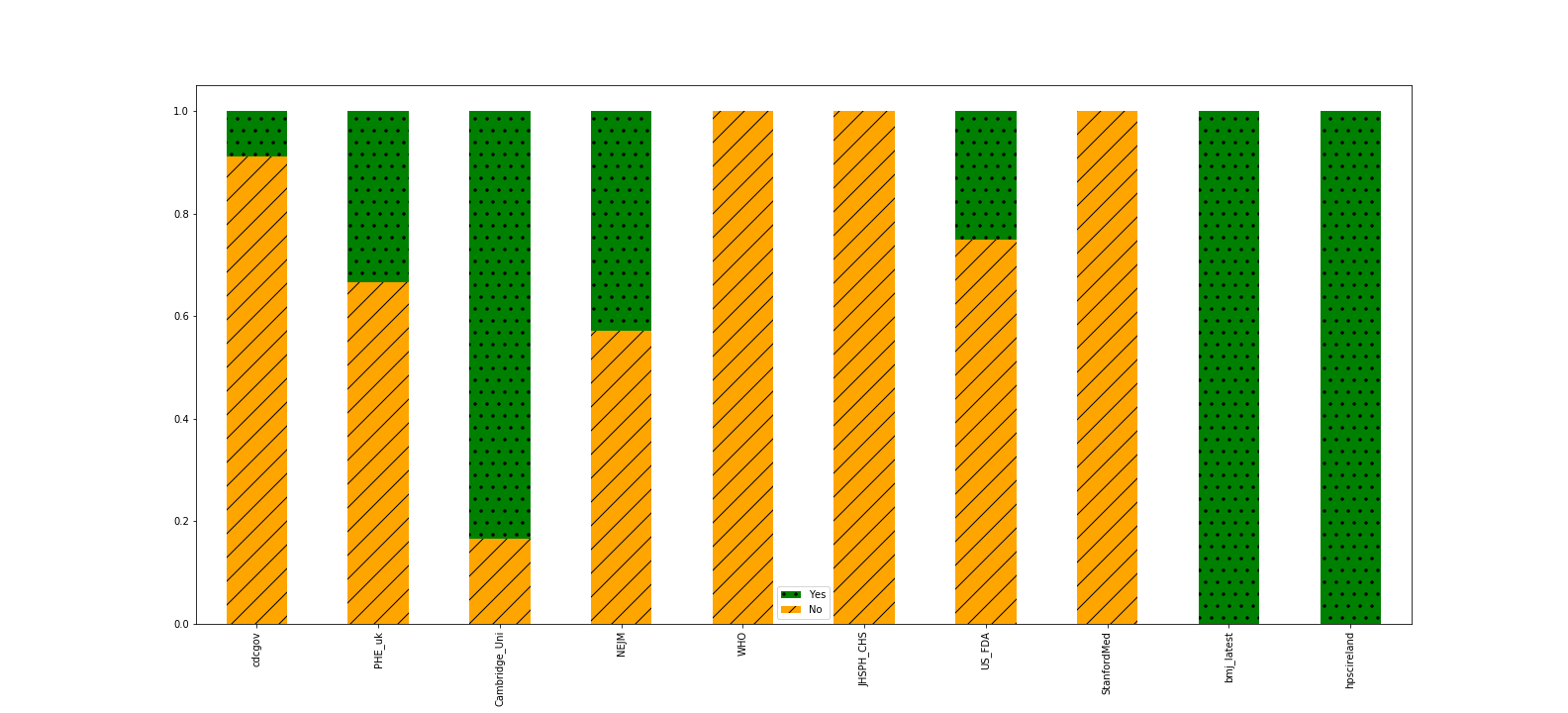}}
    \caption{Top 10 sources with belief percentages.}
    \label{fig:sourcebelief}
\end{figure}

By plotting belief relative to each scientific source, we are able to see which sources receive more trust. This could be for multiple reasons, either users lend more authority to particular sources in general, or the sources are using a better approach to relay COVID-related messages. We limit the plot in Figure \ref{fig:sourcebelief} to top 10 sources, from left to right, they are: \textit{cdcgov, PHE\_UK, Cambridge\_Uni, NEJM, WHO, JHSPH\_CHS, US\_FDA, StanfordMed, bmj\_latest, hpscireland}.

We observe that some sources such as WHO are associated with mostly disbelief, whereas a source such as Cambridge University leans toward belief. Thus we continue the investigation by looking into how the information is presented for the posts that obtained higher belief scores.

\section{What Persuasion Tactics Should be Used?}
How can we relay information related to COVID to facilitate trust and belief? We have analyzed the responses to the tweets from the scientific sources, but in this section we inspect the originally posted tweets. These are 523 tweets that we split into 70\% for training and 30\% for testing. We follow the approach outlined by \cite{iyer2019} to find sentence structures and leverage structural information to profile the messaging approach of each source. In order to obtain sentence structure we generate a syntax parse tree for each tweet using TextBlob{\footnote{\url{https://textblob.readthedocs.io/en/dev/}}}. 

\subsection{Classification of Parse Trees}
\begin{figure}[tph!]
\centerline{\includegraphics[width=\linewidth]{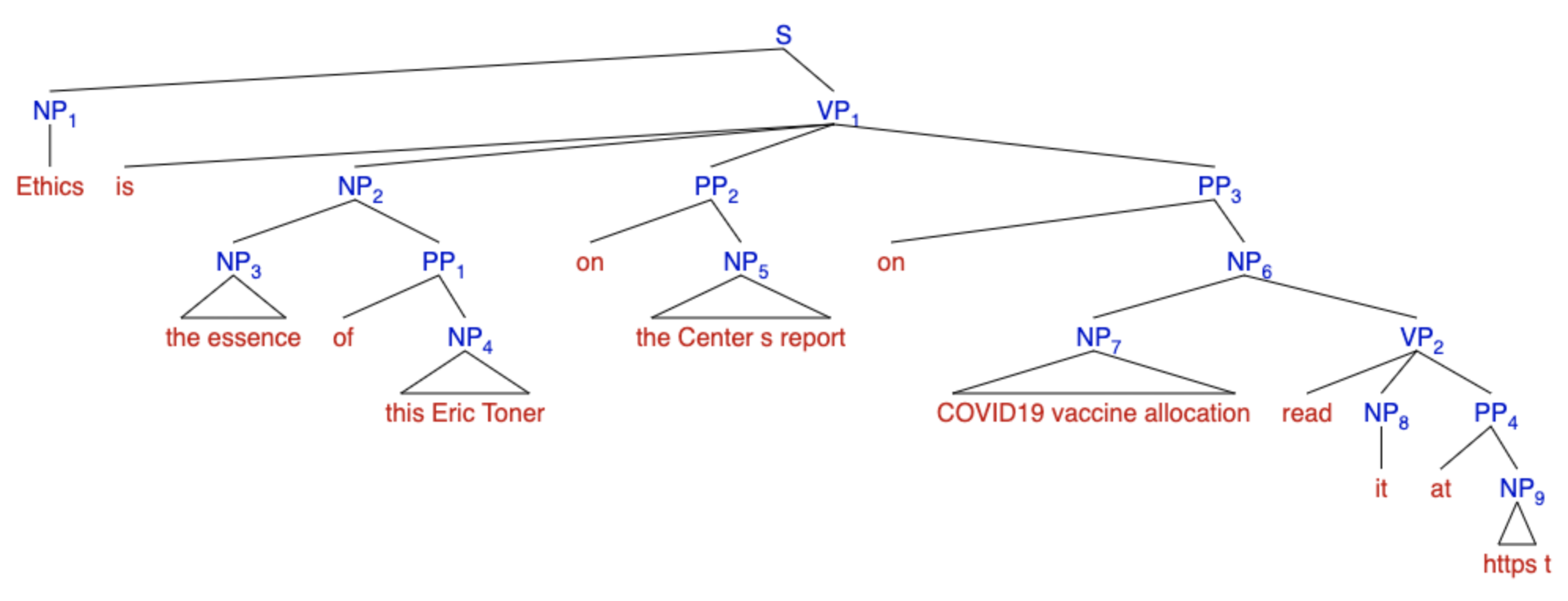}}
    \caption{Example sentence with median parse tree that receives positive belief responses.}
    \label{fig:parsetree}
\end{figure}

\begin{figure}[tph!]
\centerline{\includegraphics[width=\linewidth]{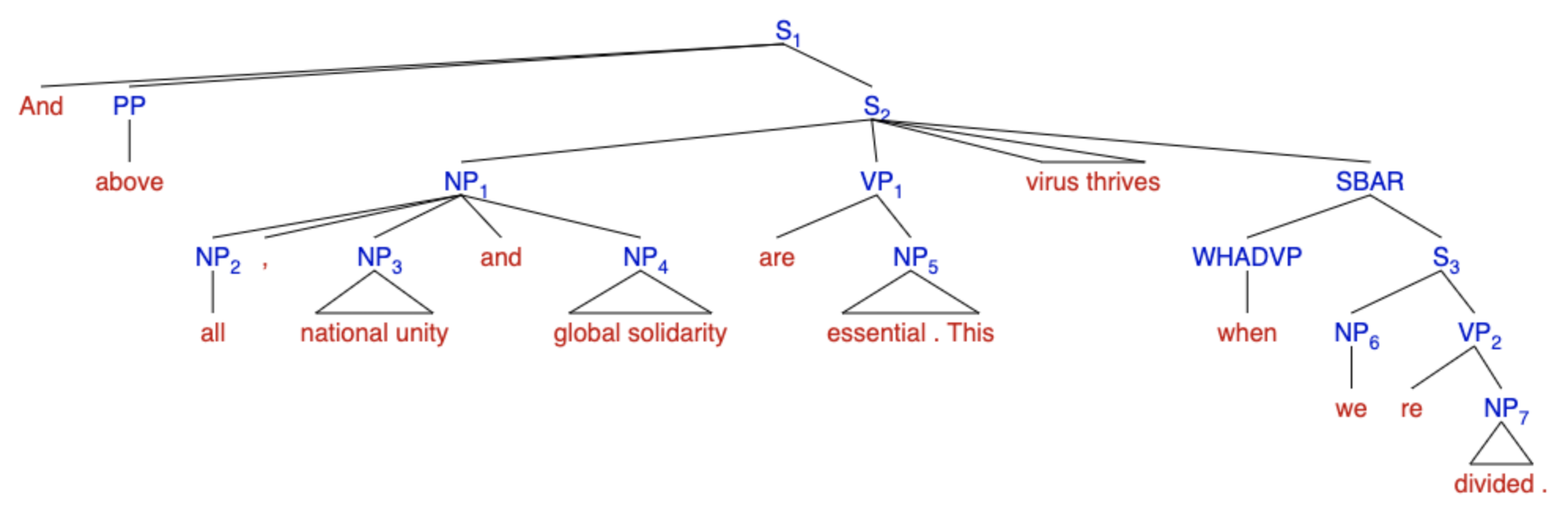}}
    \caption{Example sentence with median parse tree that receives disbelief responses.}
    \label{fig:parsetree-disbelief}
\end{figure}

Following \cite{iyer2019}, we calculate the median sentence structure for the tweets including ``COVID'' or ``corona'' which received a majority ``Yes'' responses. This is done by treating the parse trees as parse strings and calculating the pairwise Levenshtein distance between them \cite{levenshtein2009}. The median string is the one that has the minimum edit distance from the rest of the strings. An example of a median parse tree for a sentence that resulted in high belief is shown in Figure \ref{fig:parsetree}. The sentence in the figure reads: ``Ethics is the essence of this : Eric Toner on the Center’s report on \#COVID19 vaccine allocation (read it at https://t.co/ul1BZiBTtd)''. The sentiment is neutral, and the original poster is citing a secondary source. We replicate this approach for ``No'' responses, and are able to obtain a second median tree shown in Figure \ref{fig:parsetree-disbelief}. 

The sentence in Figure \ref{fig:parsetree-disbelief} reads: ``And above all, national unity and global solidarity are essential. This virus thrives when we’re divided. When we’re united, we can defeat it. -@DrTedros''. Once again, a secondary source is cited and sentiment is relatively neutral, but the message can be read as somewhat more ``preachy''. Based on the above, we calculate edit distances between each pair of tweets. Using the median belief/disbelief trees, we attempt to predict if a tweet will have a ``Yes'' response based on its sentence structure alone. We find that we can classify tweets based on the median tree that they are closest to (in terms of Levenshtein distance) to a degree of accuracy of 75.90\%. This means that syntactic structure alone can be an indicator of whether or not a tweet will be believed, and the parse tree presented in Figure \ref{fig:parsetree-disbelief} can be an outline for the syntax to be used for persuasion. Nevertheless this remains one factor in many that will be investigated in future work.

\section{Conclusions and Future Work}
In this study, we examined the impact of COVID-19 on belief as expressed by responses to scientific sources on Twitter. We were able to predict belief in responses to tweets, with a 77.19\% accuracy. By profiling conversations using hashtags, our analysis revealed a large presence of hashtags related to conspiracy theories. Our qualitative analysis indicated that topics littered with conspiracy theories exhibited little or no belief in response to messages from scientific sources. We found belief is particularly low in discussions related to COVID; almost half that of non-COVID tweets. 

We also demonstrated that the sentence structure of COVID-related tweets can be used as a signal to predict belief in responses with an accuracy of 75.9\%, thus presenting the possibility of exploring effective messaging strategies for behaviour change.

One of the biggest limitations of this project was the small amount of data, and the focus on English-language tweets. By obtaining more tweets over a longer period of time, we would be able to complete a deeper-dive on the dataset, including a time-series analysis in order to track changes in belief patterns as the virus spreads globally. Another self-imposed limitation of the data is the selection of the scientific users; the list of users would be expanded in future as more data is collected. Furthermore, investigations into persuasion tactics for more effective messaging strategies. 

\section{Acknowledgement}
This paper was prepared for informational purposes by the Artificial Intelligence Research group of JPMorgan Chase \& Co and its affiliates (``J.P. Morgan''), and is not a product of the Research Department of J.P. Morgan. J.P. Morgan makes no representation and warranty whatsoever and disclaims all liability, for the completeness, accuracy or reliability of the information contained herein. This document is not intended as investment research or investment advice, or a recommendation, offer or solicitation for the purchase or sale of any security, financial instrument, financial product or service, or to be used in any way for evaluating the merits of participating in any transaction, and shall not constitute a solicitation under any jurisdiction or to any person, if such solicitation under such jurisdiction or to such person would be unlawful.   

\begin{quote}
\begin{small}
\bibliography{aaai21}

\begin{thebibliography}{13}
\providecommand{\natexlab}[1]{#1}
\providecommand{\url}[1]{\texttt{#1}}
\providecommand{\urlprefix}{URL }
\expandafter\ifx\csname urlstyle\endcsname\relax
  \providecommand{\doi}[1]{doi:\discretionary{}{}{}#1}\else
  \providecommand{\doi}{doi:\discretionary{}{}{}\begingroup
  \urlstyle{rm}\Url}\fi

\bibitem[{Boberg et~al.(2020)Boberg, Quandt, Schatto-Eckrodt, and
  Frischlich}]{boberg2020}
Boberg, S.; Quandt, T.; Schatto-Eckrodt, T.; and Frischlich, L. 2020.
\newblock Pandemic Populism: Facebook pages of alternative news media and the
  Corona crisis - a computational content analysis.

\bibitem[{Dilrukshi, De~Zoysa, and Caldera(2013)}]{Dilrukshi2013}
Dilrukshi, I.; De~Zoysa, K.; and Caldera, A. 2013.
\newblock Twitter news classification using SVM.
\newblock 287--291.
\newblock ISBN 978-1-4673-4464-7.
\newblock \doi{10.1109/ICCSE.2013.6553926}.

\bibitem[{Ferrara(2020)}]{ferrara2020}
Ferrara, E. 2020.
\newblock What Types of COVID-19 Conspiracies are Populated by Twitter Bots?

\bibitem[{Iyer and Sycara(2019)}]{iyer2019}
Iyer, R.~R.; and Sycara, K. 2019.
\newblock An Unsupervised Domain-Independent Framework for Automated Detection
  of Persuasion Tactics in Text.

\bibitem[{Jaiswal, LoSchiavo, and Perlman(2020)}]{jaiswal_2020}
Jaiswal, J.; LoSchiavo, C.; and Perlman, D. 2020.
\newblock Disinformation, Misinformation and Inequality-Driven Mistrust in the
  Time of COVID-19: Lessons Unlearned from AIDS Denialism.
\newblock \emph{AIDS and Behavior} \doi{10.1007/s10461-020-02925-y}.

\bibitem[{{Li} et~al.(2016){Li}, {Xu}, {He}, {Deng}, and {Sun}}]{Li2016}
{Li}, J.; {Xu}, H.; {He}, X.; {Deng}, J.; and {Sun}, X. 2016.
\newblock Tweet modeling with LSTM recurrent neural networks for hashtag
  recommendation.
\newblock In \emph{2016 International Joint Conference on Neural Networks
  (IJCNN)}, 1570--1577.

\bibitem[{Miller, Vandome, and McBrewster(2009)}]{levenshtein2009}
Miller, F.~P.; Vandome, A.~F.; and McBrewster, J. 2009.
\newblock \emph{Levenshtein Distance: Information Theory, Computer Science,
  String (Computer Science), String Metric, Damerau?Levenshtein Distance, Spell
  Checker, Hamming Distance}.
\newblock Alpha Press.
\newblock ISBN 6130216904.

\bibitem[{Nourbakhsh et~al.(2019)Nourbakhsh, Liu, Li, and Shah}]{nourbakhsh}
Nourbakhsh, A.; Liu, X.; Li, Q.; and Shah, S. 2019.
\newblock Mapping the echo-chamber: detecting and characterizing partisan
  networks on Twitter.

\bibitem[{{Oscar Deho} et~al.(2018){Oscar Deho}, {William Agangiba}, {Felix
  Aryeh}, and {Jeffery Ansah}}]{Deho2018}
{Oscar Deho}, B.; {William Agangiba}, A.; {Felix Aryeh}, L.; and {Jeffery
  Ansah}, A. 2018.
\newblock Sentiment Analysis with Word Embedding.
\newblock In \emph{2018 IEEE 7th International Conference on Adaptive Science
  Technology (ICAST)}, 1--4.

\bibitem[{Pennington, Socher, and Manning(2014)}]{Pennington14glove}
Pennington, J.; Socher, R.; and Manning, C.~D. 2014.
\newblock Glove: Global vectors for word representation.
\newblock In \emph{In EMNLP}.

\bibitem[{Prochazka and Schweiger(2019)}]{prochazka2019}
Prochazka, F.; and Schweiger, W. 2019.
\newblock How to Measure Generalized Trust in News Media? An Adaptation and
  Test of Scales.
\newblock \emph{Communication Methods and Measures} 13(1): 26--42.
\newblock \doi{10.1080/19312458.2018.1506021}.
\newblock \urlprefix\url{https://doi.org/10.1080/19312458.2018.1506021}.

\bibitem[{Schild et~al.(2020)Schild, Ling, Blackburn, Stringhini, Zhang, and
  Zannettou}]{schild2020}
Schild, L.; Ling, C.; Blackburn, J.; Stringhini, G.; Zhang, Y.; and Zannettou,
  S. 2020.
\newblock "Go eat a bat, Chang!": An Early Look on the Emergence of Sinophobic
  Behavior on Web Communities in the Face of COVID-19.

\bibitem[{Sch{\"o}nfeld and Pfeffer(2019)}]{Fruchterman1991}
Sch{\"o}nfeld, M.; and Pfeffer, J. 2019.
\newblock \emph{Fruchterman/Reingold (1991): Graph Drawing by Force-Directed
  Placement}, 217--220.
\newblock Wiesbaden: Springer Fachmedien Wiesbaden.
\newblock ISBN 978-3-658-21742-6.
\newblock \doi{10.1007/978-3-658-21742-6_49}.
\newblock \urlprefix\url{https://doi.org/10.1007/978-3-658-21742-6_49}.

\end{thebibliography}
\end{small}
\end{quote}

\end{document}